\documentclass{appolb}
\usepackage{epsfig}

\begin{document}
\title{Jets in Nuclear Collisions  %
\thanks{Presented at the XXXIV International Symposium on
        Multiparticle Dynamics}%
}
\author{Ivan Vitev
\address{Los Alamos National Laboratory, Mail Stop H846, Los Alamos, 
NM 87545, USA}
}
\maketitle
\begin{abstract}

Ultra-relativistic heavy ion collisions at RHIC and the LHC 
open exciting new possibilities for jet physics studies in 
the presence of hot and dense nuclear matter. Recent theoretical 
advances in understanding the QCD multi-parton dynamics provide  
a good description of the quenching in the single and double 
inclusive high-$p_T$ hadron spectra. Measurement of the 
redistribution of the lost energy and the corresponding increase 
in the soft hadron multiplicities is the next critical step 
in elucidating the modification of the jet properties in the 
nuclear environment.

\end{abstract}
  
\section{Introduction}

Jet production~\cite{Hanson:1975fe,Sterman:1977wj} is among 
the most robust high-$Q^2$ processes, calculable within the  
pQCD factorization approach~\cite{Collins:gx}.  
In the case of heavy ion reactions the interactions of 
the hard probe with the bulk partonic matter lead 
to elastic, inelastic and coherent modifications to the cross 
section that can be systematically incorporated in the 
perturbative formalism~\cite{Vitev:2004kd}.

At large center of mass energies jet production may probe
the parton distribution functions $\phi(x,\mu_f)$ at small
momentum fractions $x \leq 0.1$. In the case of heavy ion reactions 
nuclear size enhanced power corrections generate 
dynamical parton mass and lead to nuclear shadowing~\cite{Qiu:2003vd}.
While these are relevant for low- and  moderate $Q^2$ (or  $-t$) 
processes, large $E_T$ jet production 
$Q^2$  (or $-t$) $\gg  m^2_{\rm dyn}$ remains 
unaffected~\cite{Qiu:2003vd}.

Before we investigate the consequences of medium-induced acoplanarity 
and non-Abelian bremsstrahlung some basic characteristics of 
jets in $e^+ + e^-$ and $p+p\,(\bar{p})$ collisions should be reviewed.
The virtuality $ \frac{Q}{2} \sim \sqrt{t} $ of a hard perturbative 
process is reduced  to a limiting  value $ t_0  \ll t$ via soft gluon
radiation. In the simple case of independent Poisson 
emission the induced parton multiplicities scale with average 
squared color charge of quarks and gluons $C_R  = \{C_A,C_F\}$. 
In the presence of color coherence~\cite{Dokshitzer:1978hw,Field} 
parts of the phase for soft gluon bremsstrahlung are excluded 
due to destructive interference effects. For the limiting case 
of factorizable exact double ordering~\cite{Field},  
both in terms of the lightcone momentum fractions 
$z_i$ and the virtualities $t_i$, one finds significant
corrections for the predicted soft hadron multiplicities.
Let $ \alpha_s(t) = \frac{4\pi}{\beta_0} 
\left( \ln \frac{t}{\Lambda^2_{\rm QCD}}   \right)^{-1}$ 
and $\kappa = \frac{2}{\beta_0} \ln \frac{\alpha_s(t_0)}{\alpha_s(t)} $. 
The exclusive probability  for $n$-gluon 
emission is given by  $P_n(t;t_0,z_0) = 
\left( 2 C_R \, \kappa \, \ln \frac{1}{z_0} \right)^n 
( n! n!)^{-1}$.
Standard  first  moment  evaluation  yields  an   
average  gluon multiplicity 
\begin{equation} 
\langle N_g \rangle = 
\left( \frac{\rho}{2}  \right)  \frac{I_1(\rho)}{I_0(\rho)} \;,
\quad \rho = 2 \sqrt{2 C_R \, \kappa \, \ln \frac{1}{z_0}} \;.
\label{average}
\end{equation}
It is easy to verify that in the small $z_0$ or $\alpha_s(t)$
 $\lim_{z_0\rightarrow 0\; | \;  \alpha_s  \rightarrow 0} 
\langle N_g  \rangle = \frac{\rho}{2} $.  
While    $\langle N_g \rangle$  depends on the 
choice of $z_0$ and $t_0$, assuming isospin symmetry,
$ N_{\rm ch} = \frac{2}{3} N_{\rm tot}$,  and  
local parton-hadron duality~\cite{Dokshitzer:eq} we find 
\begin{equation}
\frac{\langle  N_{\rm ch} \rangle_{\rm g-jet} }
{\langle  N_{\rm ch} \rangle_{\rm q-jet} }
\simeq \lim_{z_0\rightarrow 0\; | \; \alpha_s \rightarrow 0} 
\frac{\langle  N_g \rangle_{\rm g-jet} }
{\langle  N_g \rangle_{\rm q-jet} } = \sqrt{\frac{C_A}{C_F}} 
 = \frac{3}{2}   \;. 
\label{ratio-mult}
\end{equation}

\begin{figure}[t!]
\vspace*{0.5cm}
\includegraphics[width=2.6in,height=1.9in]{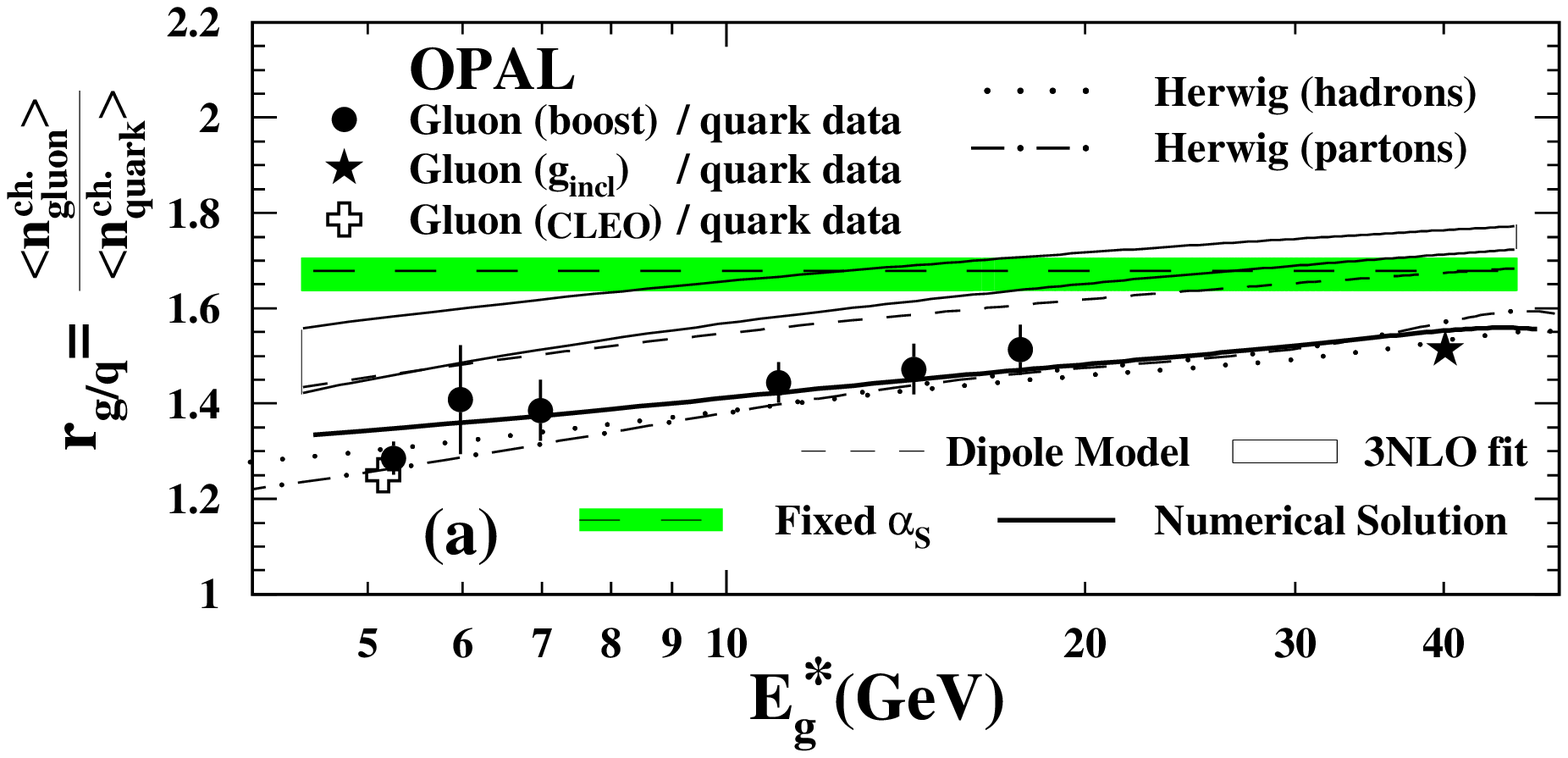}
\includegraphics[width=2.6in,height=1.7in]{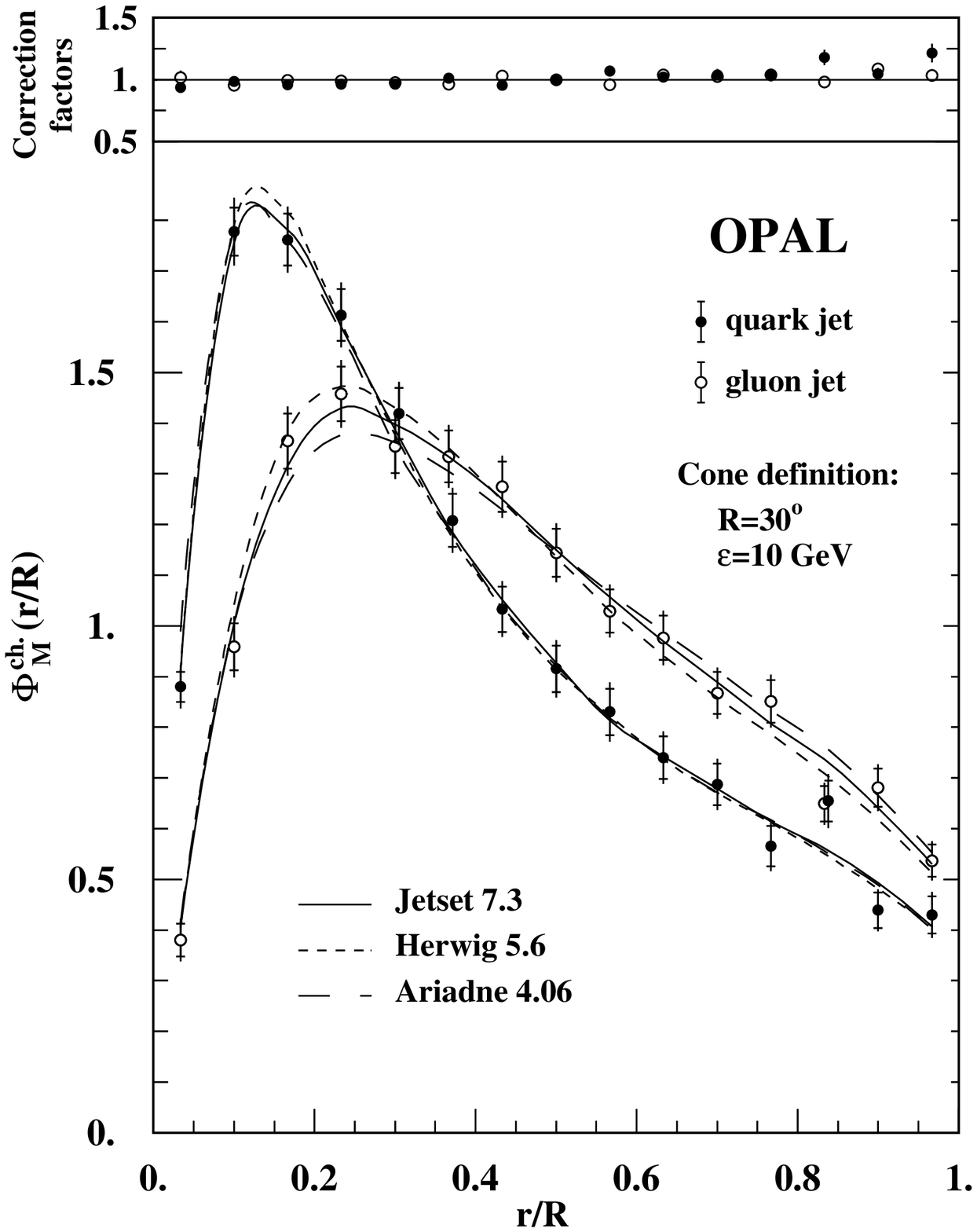} 
\vspace*{-.5cm}
\caption{ Left panel from~\cite{Abbiendi:2003gh}:
the ratio of charged hadron multiplicities associated 
with gluon and quark jets in the range E$_g^* = 5 - 40$~GeV.
Right panel from~\cite{Akers:1995im}: 
the distribution of charged hadrons in $r/R$ with
$R=30^0$. Note the broader distribution for gluon
jets.
}
\label{pp-fig}
\end{figure}

The left panel of Fig.~\ref{pp-fig} shows the ratio
of the charged hadron multiplicities  for quark and gluon jets 
measured by the OPAL collaboration~\cite{Abbiendi:2003gh}. 
Over a wide range of energies $E^*_g = p_{T\; g}$  
the experimental results fall in the range 
$\frac{\langle  N_{\rm ch} \rangle_{\rm g-jet} }
{\langle  N_{\rm ch} \rangle_{\rm q-jet} } = 1.4 - 1.6$,
which has to be compared with the analytic expectation of $1.5$ from 
Eq.~(\ref{ratio-mult}). The typical charge hadron multiplicity
for gluon jets of  $E^*_g = 10 - 20$~GeV is 
$ \langle  N_{\rm ch} \rangle_{\rm g-jet} = 6 - 10$~\cite{Abbiendi:2003gh}.

Parton broadening relative to the axis of propagation 
and the shape of the transverse momentum distributions can
be roughly estimated in the leading  double log 
approximation (LDLA). The normalized 
${\bf k}_T$ probability  from vacuum  
radiation and including Sudakov form factors~\cite{Sudakov:1954sw} 
is given by
\begin{equation}
\frac{1}{\sigma_0} \frac{d\sigma}{d {\bf k}_T^2} {\Big |}_{LDLA} 
= - 2 \,  \frac{ C_R \alpha_s}{2 \pi} \, \frac{1}{ {\bf k}_T^2 } 
\log  \frac{ {\bf k}_T^2 }{Q^2}  \, \exp \left(  
 -   \frac{ C_R \alpha_s}{2 \pi} \,  \log^2 \frac{ {\bf k}_T^2 }{Q^2} 
  \right) \; . 
\label{sudak}   
\end{equation}
The simple analytic form, Eq.~(\ref{sudak}), has definite 
shortcomings. It forgoes important kinematic constraints, 
assigns $\equiv 0$  probability to $\sum_i {\bf k}_{T\;i}=0$ 
type  configurations and thus suggests that back-to-back leading  
hadrons always  disfavor the $\Delta \phi = \pi$ topology.
Experimentally, large away-side correlations have been 
measured in $p+p$ and $d+Au$ reactions at 
$\Delta \phi = \pi$~\cite{Adler:2002tq}.

The mean transverse momentum broadening  from 
Eq.~(\ref{sudak}) reads
\begin{equation} 
 \langle {\bf k}_T^2 \rangle_{pp}  =  
\left(  1 - \frac{\pi}{\sqrt{2 C_R \alpha_s} }   
e^{\frac{\pi}{ 2 C_R \alpha_s} } \left[ 1 - 
{\rm Erf} \left( \sqrt{ \frac{\pi}{2 C_R \alpha_s}} \right)    
\right]    \right) \;Q^2  \; 
\label{mean-kT}   
\end{equation}
and is proportional to the only dimensionful scale in the 
problem $Q^2$. In the small coupling limit 
$ \langle {\bf k}_T^2 \rangle_{pp} = \frac{ C_R \alpha_s }{ \pi} \, 
(1+ {\cal O}(\alpha_s) )  \, Q^2$. 
As emphasized above, correction will likely reduce the $Q^2$ 
dependence of these estimates. Nevertheless, one still expects
a strong correlation between the acoplanarity  momentum 
projection $ \langle k_{T\, y} \rangle $~\cite{Adler:2002tq}
and the hardness of  the process. The ratio of the 
broadening and the width of the jet cone for quark and gluon 
jets is approximately given by 
\begin{equation}
\lim_{\alpha_s \rightarrow 0} \frac{\theta_g}{\theta_q} \sim 
 \sqrt{   \frac{\langle {\bf k}_T^2 \rangle_{\rm g-jet} } 
{\langle {\bf k}_T^2 \rangle_{\rm q-jet} }  } =  
\sqrt{ \frac{C_A}{C_F}     } =  \frac{3}{2} \; .
\label{broadrat}
\end{equation}
Differences in the angular distribution of hadrons in
quark and gluon jets from the OPAL experiment are shown in the 
right hand side of Fig.~\ref{pp-fig}.

\section{ Modification of the jet properties in nuclear 
collisions }

\begin{figure}[t!]
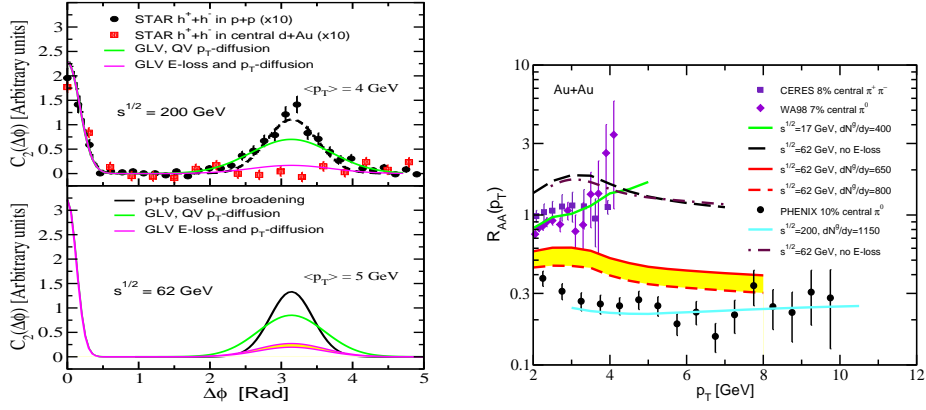

\vspace*{-.0cm}
\includegraphics[width=2.2in,height=2.1in]{corr-ISMD.eps} 
\hspace*{0.5cm}
\includegraphics[width=2.3in,height=1.8in]{Quench-62GeV.eps}
\caption{ 
Right panel: the away-side correlation function 
$C_2(\Delta \phi)$ in $p+p$ and central $Au+Au$ reactions 
with transverse 
momentum diffusion and with and without jet energy loss. 
Data is from STAR~\cite{Adler:2002tq}. 
Left panel from~\cite{Vitev:2004gn}:
predicted suppression ratio $R^{h_1}_{AA}(p_T)$  
for neutral pions at $\sqrt{s}_{NN}=17$, $62$ and  $200$~GeV.   
SPS and RHIC data~\cite{d'Enterria:2004ig} 
is shown for comparison.
}
\label{quench}
\end{figure}

In dense nuclear matter one of the anticipated modifications 
of the jet properties is the accumulation of transverse momentum 
from  elastic multi-parton interactions in addition to the 
vacuum acoplanarity discussed in Sec.~I, 
 \begin{equation}
\langle {\bf k}_T^2 \rangle_{\rm tot} = 
\langle {\bf k}_T^2 \rangle_{\rm pp} +
 \langle {\bf k}_T^2 \rangle_{\rm nucl} \; .
\label{total-kt2}
\end{equation}
$p_T-$diffusion~\cite{Qiu:2003pm}, amplified by the underlying
steep partonic slope, results in the Cronin effect observed 
in $p+A$  reactions~\cite{Vitev:2002pf}. Constraints from 
fits to low 
energy data~\cite{Cronin:1974zm} suggest that in  cold nuclear 
matter at midrapidity 
$\langle {\bf k}_T^2 \rangle \simeq 0.7$~GeV$^2$ per 
jet~\cite{Qiu:2003vd}. Such broadening is relatively small 
compared to the acoplanarity from vacuum bremsstrahlung, 
Eq.~(\ref{mean-kT}), especially in hard processes.

Significantly stronger  $p_T-$diffusion is expected in 
hot nuclear matter of initial effective gluon 
rapidity density $\frac{dN^g}{dy} \sim 1000$
and  $\rho_g(\tau) = \frac{1}{\tau A_\perp} \frac{dN^g}{dy} $, 
as shown in the left hand side of Fig.~\ref{quench}. Comparisons  
to existing data~\cite{Adler:2002tq,Miller:2004ri} on
di-hadron correlations 
$ C_2(\Delta \phi) = \frac{1}{N_{\rm trig}} 
\frac{dN^{h_1h_2}}{d \Delta \phi} $, however,   
demonstrate that this is {\em not} the dominant nuclear effect.     
Inelastic final state parton scattering, manifest in the 
multi-hadron attenuation ratio~\cite{Qiu:2003vd}     
\begin{equation}
\!  R^{(n)}_{AB}  =  \frac{d\sigma^{h_1 \cdots h_n}_{AB} / 
dy_1 \cdots dy_n d^2p_{T_1} \cdots d^2p_{T_n}} 
{\langle N^{\rm coll}_{AB} \rangle\, d\sigma^{h_1 \cdots h_n}_{NN} / 
dy_1 \cdots dy_n d^2p_{T_1} \cdots d^2p_{T_n}} \; ,
\label{multi}
\end{equation}
is the the signature difference between 
the $p+A$ and $A+A$ dynamics in the high$-p_T$ sector. 

The non-Abelian energy loss of jets can be calculated 
using the GLV approach~\cite{Gyulassy:2000er}.  
In the physical case of 1+1D  Bjorken expansion to 
first order in opacity~\cite{Gyulassy:2000gk}  
\begin{equation}
 \Delta E  \approx 
 \int dz \,  \frac{9 C_R \pi \alpha_s^3}{4} \rho^g(z) 
\, \ln \frac{2E}{\mu^2 \langle L \rangle } 
=  \frac{9  C_R \pi \alpha_s^3}{4} 
\frac{1}{A_\perp}  \frac{dN^{g}}{dy} \langle L  \rangle 
\,   \ln \frac{2E}{\mu^2 \langle L \rangle}  \; . 
\label{deltae}
\end{equation}

Current jet quenching calculations go beyond the mean 
$\Delta E$ approximation, Eq.~(\ref{deltae}), but  
assume independent Poisson medium-induced 
emission~\cite{Baier:2001yt}. The corresponding 
{\em increase} in the soft hadron multiplicities then scales 
as $\frac{C_A}{C_F}$ in contrast to the vacuum 
bremsstrahlung result, Eq.~(\ref{ratio-mult}). 
The right hand side of Fig.~\ref{quench} shows the predicted 
nuclear modification  $R^{h_1}_{AA}$
in central $Au+Au$ collisions at $\sqrt{s_{NN}} = 17, \; 62$ and 
$200$~GeV~\cite{Vitev:2002pf,Vitev:2004gn},  which is 
dominated by parton energy  loss.  The theoretical
calculation is in good agreement with the moderate- and 
high-$p_T$ dependence of the measured nuclear 
suppression~\cite{d'Enterria:2004ig}. 
Critical test of jet tomography~\cite{Vitev:2002pf} will be provided 
by the upcoming  $\sqrt{s_{NN}}=62$~GeV  pion  attenuation 
data. Parton energy loss also leads to
the suppression of the double inclusive hadron production
$R^{h_1h_2}_{AA}$ and is experimentally manifest as a 
reduction of the area  $A_{\rm Far}$ of the away-side 
correlation function $C_2(\Delta \phi)$. Such attenuation 
is $25\% - 50\%$ larger than the suppression in the single 
inclusive spectra. Numerical results are shown in the left hand side 
of Fig.~\ref{quench}.

\begin{figure}[t!]
\vspace*{-.3cm}
\includegraphics[width=2.7in,height=2.1in]{Vitev-Reappear.eps}
\hspace*{0.3cm}\includegraphics[width=2.1in,height=1.8in]{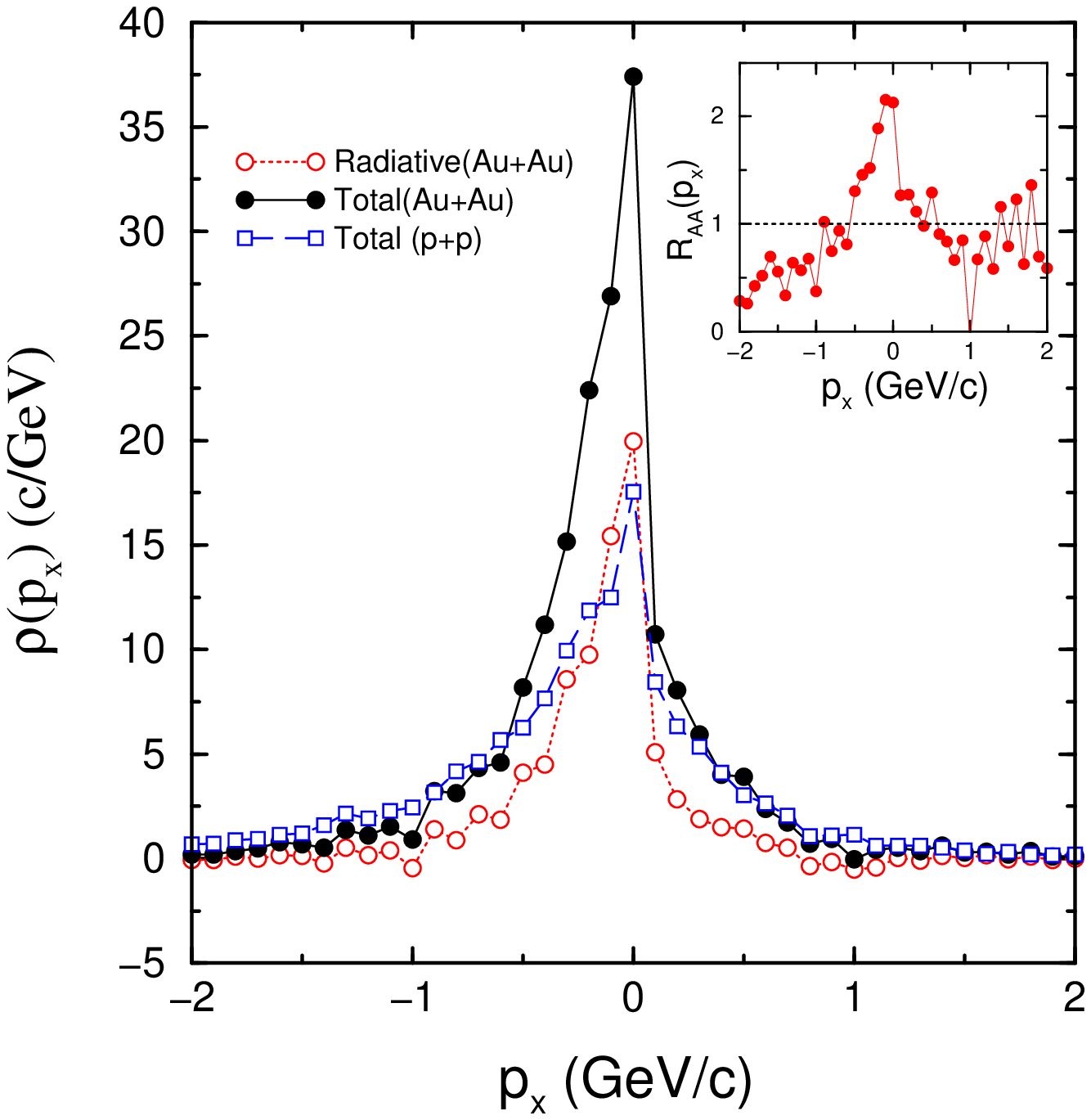} 
\caption{ Left panel from~\cite{Vitev:2002pf}: 
medium-induced partonic multiplicity as a function of the 
experimental $p_{T\, {\rm cut}}$ for energetic quark jets at RHIC and 
the LHC. Right panel from~\cite{Pal:2003zf}: momentum density 
of hadrons associated with  energetic back-to-back jets with and 
without medium-induced bremsstrahlung. Secondary rescattering leads
to gluon transverse momenta $\sim 600$~MeV. }
\label{reapp}
\end{figure}

An emerging  novel aspect  of  jet  tomography 
of  the dense  quark-gluon plasma (QGP) is the study of the 
redistribution of the lost energy, Eq.~(\ref{deltae}), 
back into the partonic system~\cite{Pal:2003zf}. With suppressed 
gluon propagation for $ \omega < \omega_{\rm  pl}$, the 
medium-induced virtuality is irradiated into  fewer harder quanta 
above the plasmon frequency~\cite{Gyulassy:2000er,Djordjevic:2003be}.  
For perfect angular acceptance, as a function of the 
experimental $p_{T \rm cut}$ for the measured hadrons  
the induced multiplicities and the total energy  recovered
in the jet are given by 
\begin{eqnarray}
\label{pt-el}
N (p_{T \rm cut} ) 
&=& 1|_{E-\Delta E \geq p_{T \rm cut} }
 +   \sum_n  n  P_n(\bar{N}_{\rm g}) \
|_{ \frac{ \Delta E}{n} \geq p_{T \rm cut}} \;,  \\
E (p_{T \rm cut} ) &=& E-\Delta E 
|_{E-\Delta E \geq p_{T \rm cut} }
 +  \frac{ \Delta E }{1 - P_0(\bar{N}_{\rm g})}  \sum_n   
 P_n(\bar{N}_{\rm g}) 
|_{\frac{\Delta E}{n} \geq p_{T \rm cut}}  \; . \qquad 
\label{pt-cuts}
\end{eqnarray}
In Eqs.~(\ref{pt-el}) and (\ref{pt-cuts}) $\bar{N}_{\rm g}$ and 
the  probability  distribution $P_n(\bar{N}_g)$  are computed 
as in~\cite{Vitev:2002pf,Baier:2001yt}. 
The left panel of Fig.~\ref{reapp} shows 
numerical estimates for $p_T = 8$ and $20$~GeV  quark jets 
at RHIC and  $p_T = 20$ and $100$~GeV  quark jets  at the LHC. 
A large part of the lost energy  reappears already at 
$p_T \simeq 1.5$~GeV  at RHIC and  $p_T \simeq 3$~GeV at the LHC.
For ideal reconstruction of the jet-related soft 
hadrons $\lim_{p_T {\rm cut } \rightarrow 0} E(p_T) = E^{\rm tot}_{\rm jet}$
and  $\lim_{p_T {\rm cut }\rightarrow 0}  N_{\rm parton}(p_{T \rm cut})
=  \bar{N}_{\rm g}(E_{\rm jet}) +1 $. The medium-induced increase 
in the parton multiplicity is $\sim 25 - 35 \%$ relative to the 
vacuum bremsstrahlung  result~\cite{Abbiendi:2003gh}.

If the radiative gluons reinteract with the QGP, their momentum
will be further degraded~\cite{Pal:2003zf}.
For complete thermalization 
$N_g({\bf r}, \Delta\tau) \approx \frac{1}{4}\Delta S =
\frac{1}{4} \frac{\Delta E({\bf r},\Delta\tau)} {T({\bf r},\tau)}$.
Numerical simulations based on a parton cascade model~\cite{Pal:2003zf} 
are shown in the right hand side of Fig.~\ref{reapp}. 
The growth of the soft multiplicity per jet is close to a factor
of two and the bremsstrahlung gluons appear at transverse momenta 
$p_T  \sim 600$~MeV.

\section{Conclusions}

The study of jets in nuclear collisions is  a natural 
extension of the calculable perturbative QCD dynamics 
to a  complex strongly interacting many-body system.  
Elastic, inelastic and coherent multiple scattering~\cite{Vitev:2004kd} 
can modify the jet and hadronic cross sections, 
the multi-hadron correlations, the energy flow of jets and 
the associated soft particle multiplicities  relative to 
measurements in baseline systems such as  
$e^+ + e^-$ and $p+p\,(\bar{p})$. 
For large $E_T$ processes it is  the medium-induced 
non-Abelian bremsstrahlung that dominates  
the observable nuclear effects. At present, the 
quenching of the single inclusive spectra and the di-hadron 
correlations is well established experimentally and understood
theoretically. The balance between the lost energy and the 
per jet growth of the soft particle production is the 
emerging novel aspect of jet tomography of dense nuclear matter.    
Preliminary results on this class of observables at RHIC 
are encouraging~\cite{Wang:2004kf} and hint at the redistribution
of the energy lost by the parent parton of the away-side jet 
into $p_T \leq 2 - 3 $~GeV hadrons. 
Improved jet reconstruction techniques for heavy ion collisions 
at RHIC and the LHC, extended $E_T$-reach and larger cross 
sections will greatly facilitate the studies of the modification 
of the energy flow and hadron multiplicities associated with 
jets in the nuclear environment.

\vspace*{.2cm}

\noindent {\bf Acknowledgments:} I would like to thank Bill Gary 
for useful discussion. This work is supported by the 
J.R.~Oppenheimer Fellowship of the Los Alamos National Laboratory 
and by the US Department of Energy.

\end{document}